\documentclass[12pt]{iopart}

\usepackage{iopams}  

\usepackage{graphicx}
\usepackage{tensor}
\usepackage{epstopdf} %for texshop on mac

\newcommand{\bra}[1]{\left\langle{#1}\right\vert}
\newcommand{\ket}[1]{\left\vert{#1}\right\rangle}
\newcommand{\ketbra}[2]{|#1\rangle \langle#2|}
\newcommand{\openone}{1 \hspace{-1.0mm}  {\bf l}}

\begin{document}

\title[]{Experimental characterization of universal one-way quantum computing}

\author{B.~A.~Bell,$^1$ M.~S.~Tame,$^2$ A.~S.~Clark,$^3$ R.~W.~Nock,$^1$ W.~J.~Wadsworth,$^4$ and J.~G.~Rarity$^1$}

\address{$^1$Centre for Communications Research, Department of Electrical and Electronic Engineering, University of Bristol, Merchant Venturers Building, Woodland Road, Bristol, BS8 1UB, UK \\ $^2$Experimental Solid State, The Blackett Laboratory, Imperial College London, Prince Consort Road, SW7 2BW, United Kingdom \\ $^3$ Centre for Ultrahigh-bandwidth Devices for Optical Systems, The Institute of Photonics and Optical Science, School of Physics, University of Sydney, NSW 2006, Australia \\ $^4$Centre for Photonics and Photonic Materials, Department of Physics, University of Bath, Claverton Down, Bath, BA2 7AY, United Kingdom}

\begin{abstract}
We report the characterization of a universal set of logic gates for one-way quantum computing using a four-photon `star' cluster state generated by fusing photons from two independent photonic crystal fibre sources. We obtain a fidelity for the cluster state of $0.66 \pm 0.01$ with respect to the ideal case. We perform quantum process tomography to completely characterize a controlled-NOT, Hadamard and T gate all on the same compact entangled resource. Together, these operations make up a universal set of gates such that arbitrary quantum logic can be efficiently constructed from combinations of them. We find process fidelities with respect to the ideal cases of $0.64 \pm 0.01$ for the CNOT, $0.67 \pm 0.03$ for the Hadamard and $0.76 \pm 0.04$ for the T gate. The characterisation of these gates enables the simulation of larger protocols and algorithms. As a basic example, we simulate a Swap gate consisting of three concatenated CNOT gates. Our work provides some pragmatic insights into the prospects for building up to a fully scalable and fault-tolerant one-way quantum computer with photons in realistic conditions.
\end{abstract}

\maketitle

%%%%%%%%%%%%%%%%%%%%%%%%%%%%%%%%%%%%%%%%
\section{Introduction}

The one-way model has radically changed perspectives of quantum computation since its introduction and development over the past few years~\cite{oneway,oneway2,develop}. In the standard circuit model, logic gates are performed on elements of a quantum register to process information, in an analogue of a classical computer~\cite{NC}. The quantum logic gates are unitary operations, and hence the entire process is reversible, up until readout of the final output of the register. In contrast, in the one-way model, a multipartite entangled state known as a cluster or graph state is used as a resource to run a quantum computation which is carried out by a sequence of single-qubit measurements, which progressively collapse the wavefunction and remove the entanglement of the state, an inherently irreversible process. In addition to providing insight into the requirements for quantum computing and emphasising the importance of entanglement as a resource, applying this approach to linear optical quantum computing with photonic cluster states has led to increased flexibility and a significant reduction in the experimental resources required compared to previous schemes~\cite{onewayta,onewaytb,onewaytc,onewaye}. For quantum computing in general, a quantum controlled-NOT (CNOT) gate and arbitrary single-qubit rotations represent fundamental building blocks: taken together one can perform universal quantum computation~\cite{NC}. It has been shown that given a large enough cluster state, both CNOT gates and arbitrary rotations can be performed and combined in order to realise any quantum logic operation desired~\cite{oneway}. Very recently an individual CNOT gate was experimentally demonstrated using a six-qubit photonic cluster state~\cite{Pan}. Here, in contrast to this and earlier studies we experimentally demonstrate and fully characterize a complete set of building blocks for universal one-way quantum computing using a compact four-qubit photonic cluster state, which itself forms a fundamental building block for growing an arbitrarily large cluster state resource. The universal gates we characterise are highly efficient due to their compact nature and therefore less subject to noise and imperfections~\cite{onewaytd}. In the long-term, our method may therefore have greater potential than previous attempts for achieving the error thresholds required for fault-tolerant one-way quantum computing~\cite{develop}.

In this work, we make use of high-brightness photonic crystal fibre (PCF) sources of entangled photon pairs~\cite{Fulconis3, Clark} and a postselected fusion gate to convert the two pairs into a four-photon `star' cluster state. Fusion gates can be used to build larger cluster states in a scalable fashion beginning from a source of smaller entangled states, such as Bell states~\cite{onewaytb,onewaytc}. We have previously reported an experimental characterisation of the fusion process, which gives important information about how the quality of resource states will scale with multiple fusions~\cite{Bell}. Here, we use a fused four-photon star cluster state to demonstrate a set of gates which is universal for one-way quantum computation, {\it ie.} a CNOT, Hadamard, and T gate, and perform quantum process tomography on these operations. Unlike other experiments to generate photonic cluster states using bulk optics crystal-based sources~\cite{onewaye, clusterexperiments}, our state is made up of photons with two different wavelengths. We show that it is still possible to build larger cluster states using fusion gates in this unusual setting. While the fidelity of our state is lower than recent bulk optics implementations have achieved, largely due to imperfections in the PCFs compared to more established sources, we expect this to improve in future work, where the compactness, efficiency, and ease of integration into all-fibre experiments would become key advantages for applications in quantum networking and communication. Furthermore, while one-way quantum logic gates have been demonstrated in schemes using hyper-entanglement, or multiple degrees of freedom of each photon to encode additonal qubits~\cite{multidegfreedom} (using fewer photons to encode more information with high fidelities), the complexity of these techniques is likely to limit their scalability, or at least the number of qubits encoded per photon. Thus, different to previous studies, here we provide a detailed investigation into the effects of realistic experimental conditions in one-way quantum computing and the accumulation of errors  important for considering the scaling up to larger resources for carrying out more complex protocols. With our results and described techniques it is possible to simulate and predict the performance of any quantum logic operation using photonic cluster state computation, and as a basic example we simulate a Swap gate.

%%%%%%%%%%%%%%%%%%%%%%%%%%%%%%%%%%%%%%%%
\section{Experimental setup}
\begin{figure*}[t]
\begin{minipage}{16cm}
\hspace*{0cm}\includegraphics[width=15.7cm]{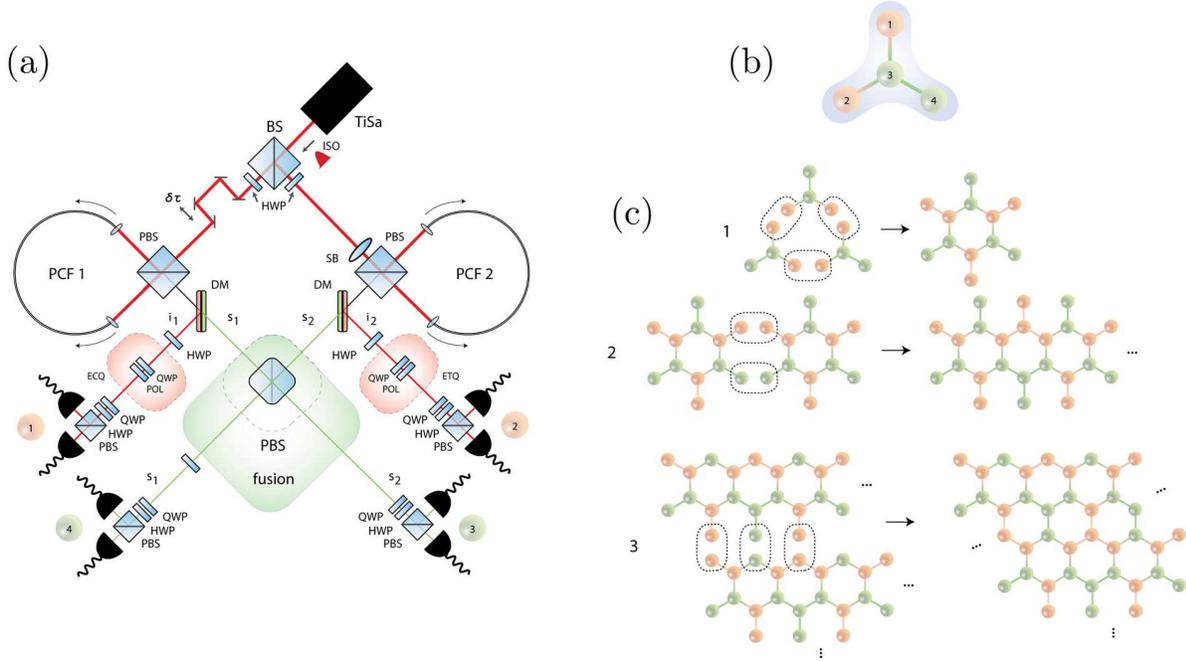}
\end{minipage}
\caption{(a)~Experimental setup. Two photonic crystal fibre sources (PCF 1 and 2) are used to produce pairs of Bell states which are fused using a polarising beamsplitter (PBS) to generate the entangled four-photon resource. (b)~The `star' cluster state generated using (a), with two green qubits (signal photons at $\lambda=625~nm$) and two red qubits (idler photons at $\lambda=860~nm$). The edges correspond to controlled-phase operations, $CZ={\rm diag}\{ 1,1,1,-1 \}$, applied to qubits (representing the vertices) initialized in the state $|+ \rangle = (\ket{0}+\ket{1})/\sqrt{2}$.  (c)~Building up a universal resource. Due to the generated cluster state being made from photons with two different wavelengths, standard methods for generating a larger resource for universal quantum computation cannot be used. Here, we show steps that can be taken in order to build up a larger universal resource using the fusion operation characterised in Ref.~\cite{Bell} (dashed rings) and adapting the techniques of Ref.~\cite{onewaytc}. The final resource has an unusual hexagonal structure, however, it can be converted to a square 2D cluster state by using simple local operations~\cite{Nest}, thus confirming it as a universal resource. However, in general, this last step is not necessary for implementing one-way quantum computing (as we show later), with the hexagonal lattice having a natural intrinsic robustness to noise due to the reduced vertex degree of 3~\cite{Dur}.}
\label{setup}
\end{figure*}

The experimental setup is shown in Figure~\ref{setup}~(a). A Ti:Sapphire laser producing picosecond pulses at $720~nm$ is split and used to pump two separate birefringent PCF sources, producing correlated pairs of photons via spontaneous four-wave mixing (FWM) at a signal and idler wavelength of $625~nm$ and $860~nm$ respectively~\cite{Fulconis3}. An intrinsically pure state phasematching scheme is used such that the signal and idler photons are generated without spectral or temporal correlations~\cite{Clark,Halder}. This allows good quality quantum interference to take place between independent sources without the need for narrow spectral filtering, enabling high lumped detection efficiencies of up to $25\%$ including collection, filtering, and detection. Each source is in a Sagnac loop configuration around a polarizing beamsplitter (PBS), such that the fibre is pumped in both directions, with the birefringent axes of the fibre oriented so that one direction produces horizontally polarized pairs of photons  in the state $\ket{H}_s\ket{H}_i$, while the other direction produces vertically polarized photons $\ket{V}_s\ket{V}_i$. When the two directions are recombined at the PBS, the resulting state outside the loop is entangled in polarization:
\begin{equation}
\ket{H}_s\ket{H}_i+e^{i\theta}\ket{V}_s\ket{V}_i.
\end{equation} 
A Soleil Babinet birefringent compensator (SB) placed before the loop is used to set the phase $\theta$ to be zero, so that the Bell state $\ket{\phi^{+}}$ is produced. This method of generating entanglement has previously been demonstrated both with fibre~\cite{Li} and bulk crystal sources~\cite{Fedrizzi}, and benefits from the high stability of a Sagnac interferometer, where the two paths are overlapping in space so that their relative length (and phase) is not affected by thermal fluctuations or vibrations.

When both entangled sources simultaneously produce a photon pair, the combined state is
\begin{equation}
\hspace*{-1cm} \frac{1}{\sqrt{2}}(\ket{H}_{s_1}\ket{H}_{i_1}+e^{i\theta_1}\ket{V}_{s_1}\ket{V}_{i_1})\otimes \frac{1}{\sqrt{2}}(\ket{H}_{s_2}\ket{H}_{i_2}+e^{i\theta_2}\ket{V}_{s_2}\ket{V}_{i_2}).
\end{equation}
Dichroic mirrors (DM) are then used to separate the two wavelengths from each source. A tunable filter window set to $\sim 4nm$ bandwidth at $860nm$ is applied to the idler - while narrow filtering is not necessary this helps to suppress Raman background and any remaining pump light without cutting into the idler spectrum and reducing the collection efficiency. The idler modes are then collected into single-mode fibres, while the signal modes are overlapped at a PBS. If the signal photons are indistinguishable in all degrees of freedom, this will apply a fusion operation $[~\ketbra{HH}{HH}+\ketbra{VV}{VV}~]_{s_1s_2}$ to the state. This is essentially a parity measurement~\cite{Bell,Parity}, which leaves the state unchanged if the two signals are of the same polarization. However, if they are of different polarizations they will exit the PBS in the same mode and a four-fold coincidence detection across the four modes is no longer possible, so that these cases are rejected. This generation scheme relies on postselection, in that the success is conditional on the detection of the four photons and hence it is not currently scalable. This can be rectified by the addition of a quantum non-demolition (QND) measurement of the photon number in the modes after the fusion. While currently technically challenging, QND measurements are in principle possible, for instance as shown in~\cite{Imoto} and recently experimentally demonstrated~\cite{Guerlin}. In our scheme the fusion operation succeeds with $50\%$ probability, producing a four photon $GHZ$ state~\cite{GHZ}
\begin{equation}
\frac{1}{\sqrt{2}}(\ket{H}_{s_{1}}\ket{H}_{i_{1}}\ket{H}_{s_{2}}\ket{H}_{i_{2}}+e^{i(\theta_1+\theta_2)}\ket{V}_{s_{1}}\ket{V}_{i_{1}}\ket{V}_{s_{2}}\ket{V}_{i_{2}}),
\end{equation}
where the phase $\theta_1=-\theta_2$ is adjusted by the SB located at the entrance to PCF 2, as shown in Figure~\ref{setup}~(a). The signal photons then pass through wide $40nm$ bandwidth filters and are coupled into single-mode fibres, before all four photons are sent to another free-space section. Half waveplates (HWPs) set at $45^{\circ}$ apply rotations to the polarization state in three of the modes (i1, i2, and s1). These local rotations turn the wavefunction into that of the star cluster state shown in Figure~\ref{setup}~(b):
\begin{equation}
\ket{\psi_{star}}=\frac{1}{\sqrt{2}}(\ket{+}_{1}\ket{+}_{2}\ket{H}_{3}\ket{+}_{4}+\ket{-}_{1}\ket{-}_{2}\ket{V}_{3}\ket{-}_{4}),
\end{equation}
where $\ket{+}=\frac{1}{\sqrt{2}}(\ket{H}+\ket{V})$ and  $\ket{-}=\frac{1}{\sqrt{2}}(\ket{H}-\ket{V})$. Following the generation of this state, additional waveplates and polarizers are used in modes 1 and 2 to encode different inputs into the logic gates, labelled ECQ and ETQ (Encode Control Qubit and Encode Target Qubit) in Figure~\ref{setup}~(a), as explained in section 4.1. Finally, each mode contains an analysis section made up of a quarter waveplate (QWP), HWP and a PBS, allowing measurements in the Pauli $\sigma_x$, $\sigma_y$, and $\sigma_z$ bases~\cite{James}. The photons are detected by an avalanche photodiode at each PBS output, and an 8 input coincidence counter (Qumet Technologies, MT-30A~\cite{Nock}) simultaneously monitors for the 16 possible four-fold detections corresponding to one photon in each mode. Coincidence events where 5 or more detections occur simultaneously are rejected, suppressing background due to multi-pair emission or Raman photons from the PCF. Each source was operated at a coincidence-to-accidental ratio of $\sim 25$, resulting in two-fold coincidence count rates of around 13,000 per second. Individual interference visibilities before fusion were $\sim96\%$ in the $\{ \ket{H}, \ket{V}\}$ basis and $\sim90\%$ in the $\{ \ket{+}, \ket{-}\}$ basis.

%%%%%%%%%%%%%%%%%%%%%%%%%%%%%%%%%%
\section{Cluster state characterisation}

\begin{figure*}[t]
\begin{minipage}{13cm}
\hspace*{2.5cm}\includegraphics[width=13cm]{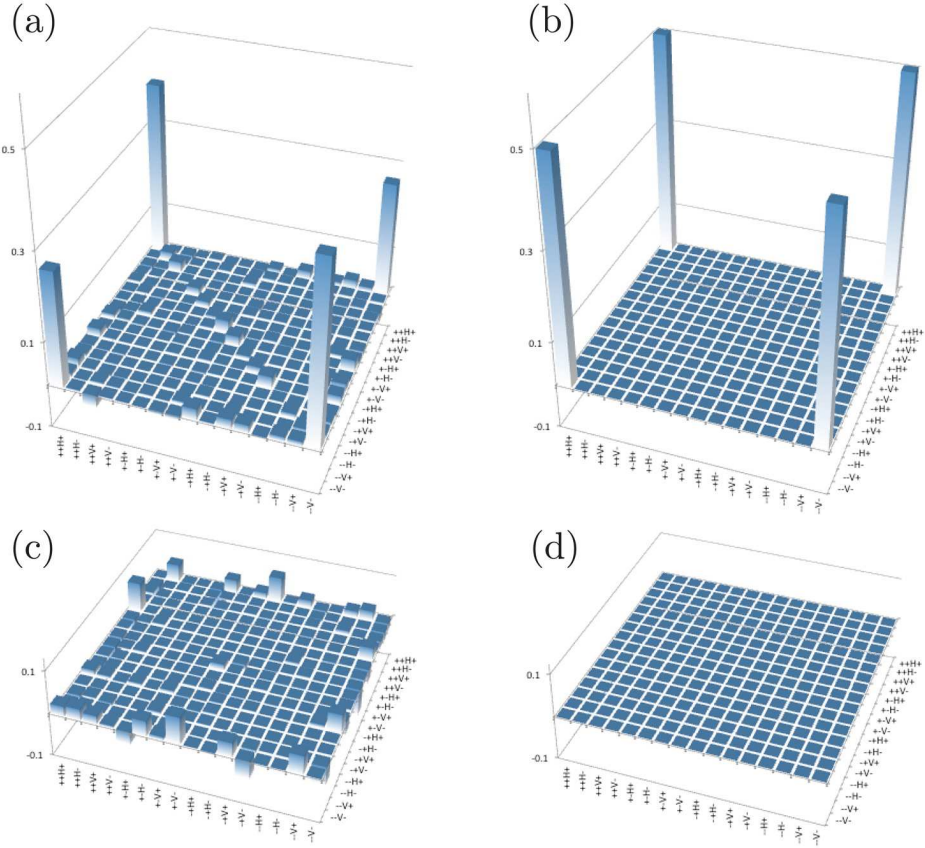}
\end{minipage}
\caption{Experimental and ideal density matrix of the star cluster state. (a)~Experimental real part. (b)~Ideal real part. (c)~Experimental imaginary part. (d)~Ideal imaginary part. For simplicity the density matrix is shown in a rotated basis. The generated state gives a fidelity of $0.66\pm 0.01$ to the ideal case.}
\label{densitymatrix}
\end{figure*}
To characterise the state generated and compare it to the ideal cluster state, we carried out state tomography, analysing the state in all combinations of three measurement bases for each photon: $\{ \ket{H}, \ket{V}\}$, $\{ \ket{+}, \ket{-}\}$, and $\{ \ket{R}, \ket{L}\}$. $\ket{R}$ and $\ket{L}$ are the right and left circular polarizations, with $\ket{R}=\frac{1}{\sqrt{2}}(\ket{H}-i\ket{V})$ and $\ket{L}=\frac{1}{\sqrt{2}}(\ket{H}+i\ket{V})$. For this, all waveplates and polarizers were removed from the encoding part of the setup. The resulting data set contains $81$ measurement bases, each of which contains 16 count rates corresponding to different measurement outcomes. These 1296 values allow the reconstruction of the density matrix $\rho_{exp}$ shown in Figure~\ref{densitymatrix}~\cite{James}. The total count rate was on average $\sim1$ coincidence per second and an integration time of 10 mins used per measurement basis. Comparing $\rho_{exp}$ to the ideal cluster state $\ket{\psi_{star}}$, we find a fidelity of $F =\bra{\psi_{star}}\rho_{exp}\ket{\psi_{star}}= 0.66 \pm 0.01$, well above the threshold of 0.5 to prove genuine four-party entanglement~\cite{GME}. The error has been calculated using a Monte Carlo method with Poissonian noise on the count statistics~\cite{James}. Examining the diagonal of the real part of $\rho_{exp}$ in Figure~\ref{densitymatrix}(a), we see that the state mainly consists of the two components expected, $\ket{++H+}$ and $\ket{--V-}$, but there are significant contributions from other components which reduce the fidelity. These background terms mainly result from multi-photon emission, Raman emission, and imperfections in the polarization optics allowing unwanted polarizations to leak through. However, the main imperfection in $\rho_{exp}$ is seen in the off-diagonal elements which indicate coherency between the $\ket{++H+}$ and $\ket{--V-}$ components. These are smaller than the ideal case, indicating some dephasing has taken place during the fusion. This is largely caused by distinguishability between the signal photons from independent sources, due to imperfect overlap of their spectral and temporal modes at the PBS, and being detected in a non-pure spectral state~\cite{Bell}. Although from theory we expect to achieve a good level of purity, inhomogeneity in the PCFs can reduce this as described in \cite{Cui}. Despite this, we show that the fidelity is sufficiently high to demonstrate the desired one-way quantum logic gates. In the future this problem is likely to be overcome either by improvement in PCF fabrication or by the use of cavity coupled pair-photon sources \cite{Zhang}.

%%%%%%%%%%%%%%%%%%%%%%%%%%%%%%%%%%
\section{Universal set of quantum logic gates}
 
\begin{figure}[t]
\centering
\includegraphics[width=15cm]{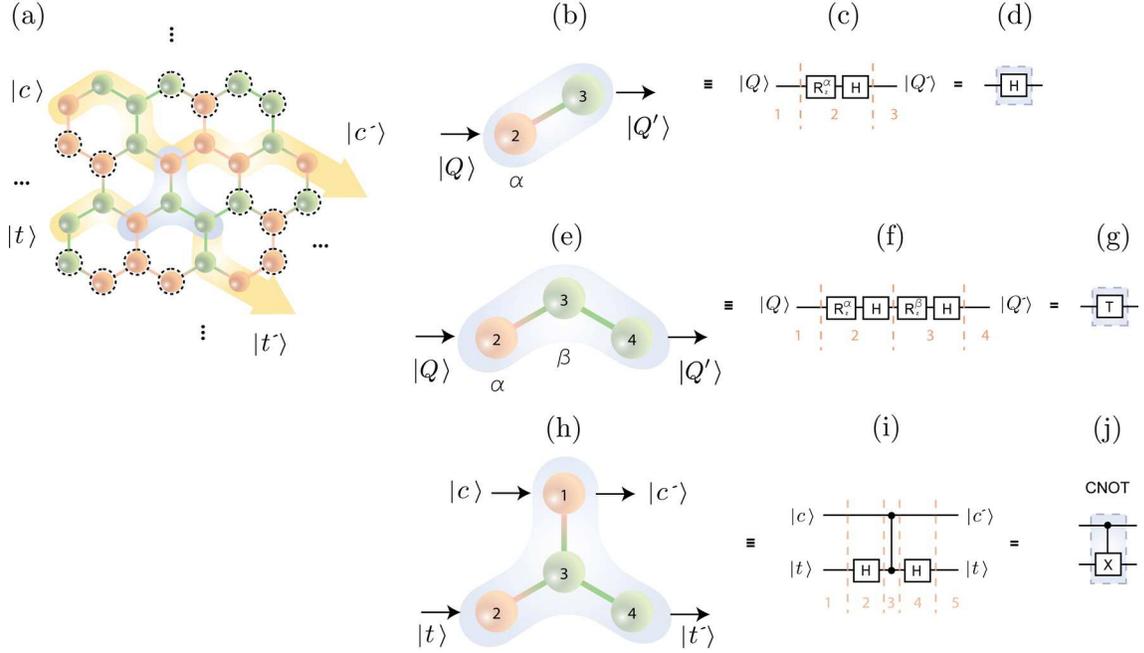}
\caption{(a)~General setting of information flow on the hexagonal lattice (yellow) and the star cluster state highlighted (blue). Here, a dotted circle denotes the qubit is removed from the resource by measuring it in the $\sigma_z$ basis. One-way implementations of (b) the {\sf H} gate, (e) the {\sf T} gate and (h) the ${\sf CNOT}$ gate on the star cluster state. Panels (c), (f) and (i) show the equivalent quantum circuits.}
\label{gates}
\end{figure}

In Figure~\ref{gates}~(a) the general setting for a star cluster state embedded within a hexagonal lattice resource is shown and the three operations required to demonstrate universal quantum logic are identified in Figures~\ref{gates}~(b)-(j). The first gate is a Hadamard rotation, {\sf H} - a single qubit operation written as the unitary matrix
\begin{equation}
{\sf H}=\frac{1}{\sqrt{2}}\left(\begin{array}{lr} 1 & 1 \\ 1 & -1\end{array}\right).
\end{equation}
It acts on the computational eigenstates such that $\ket{H} \rightarrow \ket{+}$ and $\ket{V} \rightarrow \ket{-}$. Here we use the representation $\ket{0} \equiv \ket{H}$ and $\ket{1} \equiv \ket{V}$ for the single-qubit computational basis. In the one-way model this is implemented on a logical qubit $\ket{Q}$, encoded on physical qubit 2, as shown in Figure~\ref{gates}~(b), with the gate applied by measuring physical qubit 2 in the $B(\alpha):=\{ \ket{\alpha_+}, \ket{\alpha_-}\}$ basis, where $\ket{\alpha_\pm}=(\ket{H} \pm e^{i \alpha}\ket{V})/\sqrt{2}$ and $\alpha=0$ is chosen. The logical output state $\ket{Q'}={\sf H}\ket{Q}$ is then left on physical qubit 3, up to a \emph{byproduct} operator $\sigma_x^{s_2}$, where $s_2 \in \{0,1\}$ corresponds to the outcome of the measurement on qubit 2. Pauli byproduct operators such as $\sigma_x$ are propagated through a one-way computation until the end, or compensated for during successive measurements~\cite{oneway}.

The second gate is a ${\sf T}$ gate - a single qubit rotation of $45^{\circ}$ around the $z$-axis of the Bloch sphere, written as the matrix
\begin{equation}
{\sf T}=\left(\begin{array}{lc} 1 & 0 \\ 0 & e^{i\pi /4}\end{array}\right).
\end{equation}
Combined with the Hadamard this makes up a universal gate set for efficiently constructing any single-qubit rotations~\cite{NC}. The one-way implementation is shown in Figure~\ref{gates}~(e), with qubit 2 measured in the $B(-\pi/4)$ basis and qubit 3 in the $B(0)$ basis, leaving the logical output qubit $\ket{Q'}={\sf HHT}\ket{Q}={\sf T} \ket{Q}$ on qubit 4, up to a byproduct $\sigma_x^{s_3}\sigma_z^{s_2}$. Note that this is a non-trivial gate, in the sense that the outcome of the measurement on qubit 2 determines the basis in which to measure qubit 3: $B(0)$ for $s_2=0$ or $B(\pi)$ for $s_2=1$. This is an example of adaptive measurements in the one-way model~\cite{oneway}, which impose a temporal ordering on the measurements of qubits for carrying out quantum computation~\cite{Raussendorf}. 
 
The third and final gate is the {\sf CNOT} gate, a well known example of a two-qubit gate, where a bit flip is applied to a target qubit $\ket{t}$ dependent on the state of a control qubit $\ket{c}$. It is described by the matrix
\begin{equation}
{\sf CNOT}=\left(\begin{array}{cccc} 1 & 0 & 0 & 0 \\ 0 & 1 & 0 & 0 \\ 0 & 0 & 0 & 1 \\ 0 & 0 & 1 & 0 \end{array}\right).
\end{equation}
In the circuit model this can be decomposed into two Hadamard gates applied to the target qubit before and after a controlled-phase gate $CZ:={\rm diag} \{ 1,1,1,-1 \}$, as shown in Figure~\ref{gates}~(i). Figure~\ref{gates}~(h) shows the one-way implementation, with horizontal links representing Hadamard gates and a vertical link between control and target qubits for the controlled-phase gate. The {\sf CNOT} gate is realised when qubits 2 and 3 are measured in the $B(0)$ basis, which leaves the output ${\sf CNOT}\ket{c}\ket{t}$ on the state of the remaining qubits 1 and 4, up to a local byproduct $\sigma_z^{s_2}\otimes\sigma_x^{s_3}\sigma_z^{s_2}$.

\subsection{{\sf H} gate characterisation}
\begin{figure*}[b]
\begin{minipage}{13cm}
\hspace*{3.5cm}\includegraphics[width=11cm]{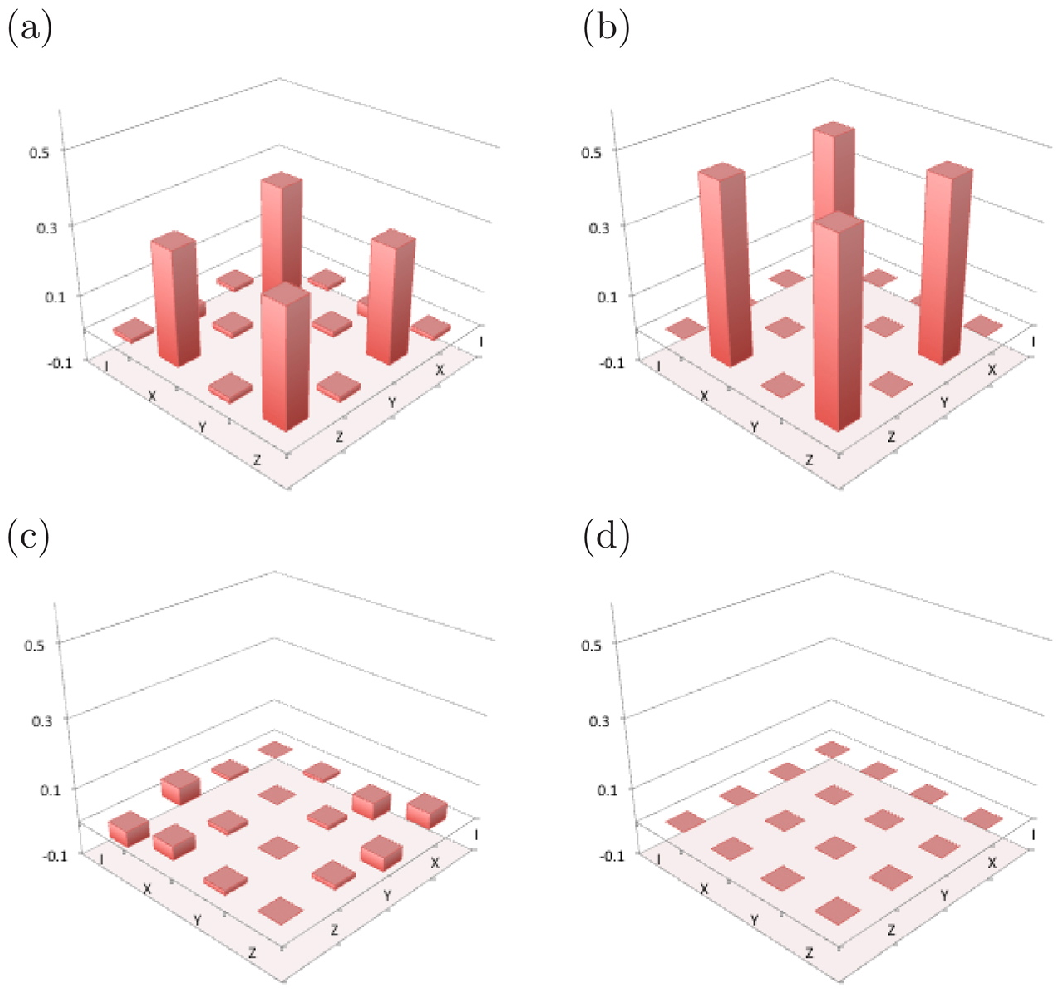}
\end{minipage}
\caption{${\sf H}$ gate $\chi$ matrix. (a) Experimental real part. (b) Ideal real part. (c) Experimental imaginary part. (d) Ideal imaginary part.}
\label{hadamard}
\end{figure*}

\begin{figure*}[b]
\begin{minipage}{13cm}
\hspace*{3.5cm}\includegraphics[width=11cm]{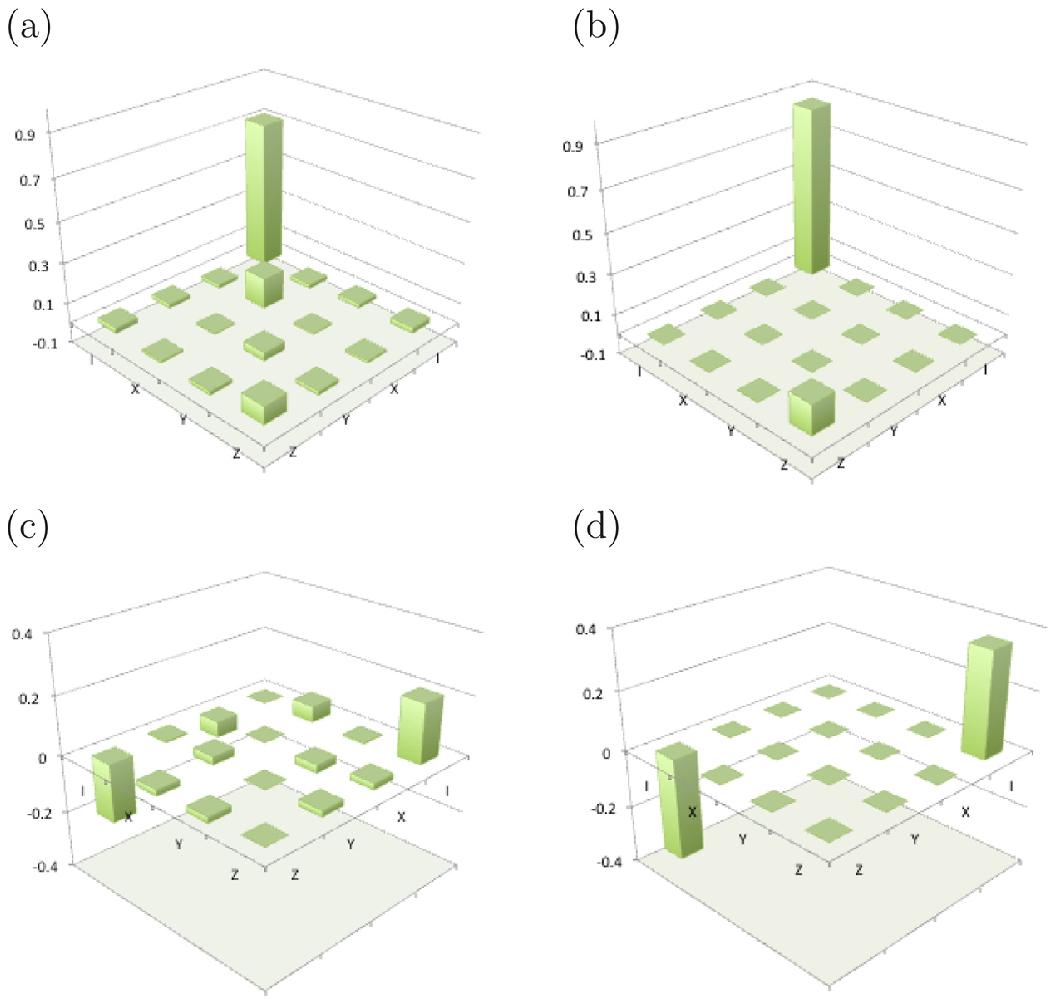}
\end{minipage}
\caption{${\sf T}$ gate $\chi$ matrix. (a) Experimental real part. (b) Ideal real part. (c) Experimental imaginary part. (d) Ideal imaginary part.}
\label{tgate}
\end{figure*}

To fully characterise the {\sf H} gate, we probe it with four input states, $\ket{H}$,  $\ket{V}$,  $\ket{+}$, and  $\ket{L}$, and carry out state tomography on the output of each. The information about the output states is then used to reconstruct the gate in the form of a quantum channel, as described below. The gate requires only two of the four physical qubits, so qubits 1 and 4 are removed from the cluster by measurements in the $\sigma_z$ basis~\cite{oneway}. The encoding is carried out by local operations on qubit 2. By default, $\ket{+}$ is encoded and therefore it requires no modification. To encode $\ket{L}$, a QWP is added with its fast axis horizontal, applying the rotation ${\rm diag}\{1, e^{i\pi /2}\}$. To encode $\ket{H}$ and $\ket{V}$, horizontal and vertical polarizers are added respectively. Note that the encoding is carried out after the entanglement is generated rather than before, so that not all input states of the qubit can be reached with unitary rotations of the physical qubit 2. Hence the use of non-unitary operations to polarize the state for  $\ket{H}$ and $\ket{V}$ inputs, which reduce the overall count rate by $50\%$. For these cases the integration time was increased correspondingly. For the measurement outcomes of all gates, we discuss in detail the case when $s_i=0,~\forall i$. Similar experimental results were obtained for the other outcomes. For the {\sf H} gate, once state tomography has been carried out on qubit 3 for each probe state, the gate can be reconstructed in the form of a channel, $ \$_g: \rho \to \rho'=\sum_{m,n}\chi_{mn}E_m \rho E_n^\dag$, where $\rho$ is any state and the $E_{m, n}$ form a complete set of Kraus operators~\cite{NC}. The experimental and ideal $\chi$ matrices for the Hadamard gate are shown in Figure~\ref{hadamard}. To ensure that the reconstructed $\chi$ matrix represents a physical process, we use a maximum likelihood technique to find a positive, Hermitian matrix that is a closest fit with the experimental data in a least-squares sense~\cite{LS} and subject to additional constraints to ensure it represents a trace-preserving process~\cite{TP}. The process fidelity $F=Tr(\chi_{ideal}\chi_{exp})/Tr(\chi_{ideal})Tr(\chi_{exp})$ is found to be $F_{\sf H}=0.67 \pm0.03$, consistent with the cluster state fidelity. The error is calculated using a Monte-Carlo approach, as described for state tomography. The basis used for the Kraus operators is $\{ \openone, \sigma_x,\sigma_y,\sigma_z \}$. Using the approach given in Ref.~\cite{Weinhold} it is possible to place a lower bound for the error probability of the Hadamard gate for the purposes of benchmarking it against fault-tolerant thresholds for scalable quantum computing. For a generalised error model~\cite{Aliferis} we have a lower bound for the minimum error-probability per gate as $\epsilon^*=1-F_{\sf H}$. Thus for the Hadamard gate we have an error-probability per gate of $\epsilon > 0.33 \pm 0.03$. This is clearly far from any fault-tolerant threshold ($\sim10^{-2}$ to $10^{-5}$~\cite{develop,Weinhold}) and therefore much improvement in the quality of the entangled resource is required.

\subsection{${\sf T}$ gate characterisation}

\begin{figure*}[b]
\begin{minipage}{13cm}
\hspace*{2.5cm}\includegraphics[width=13cm]{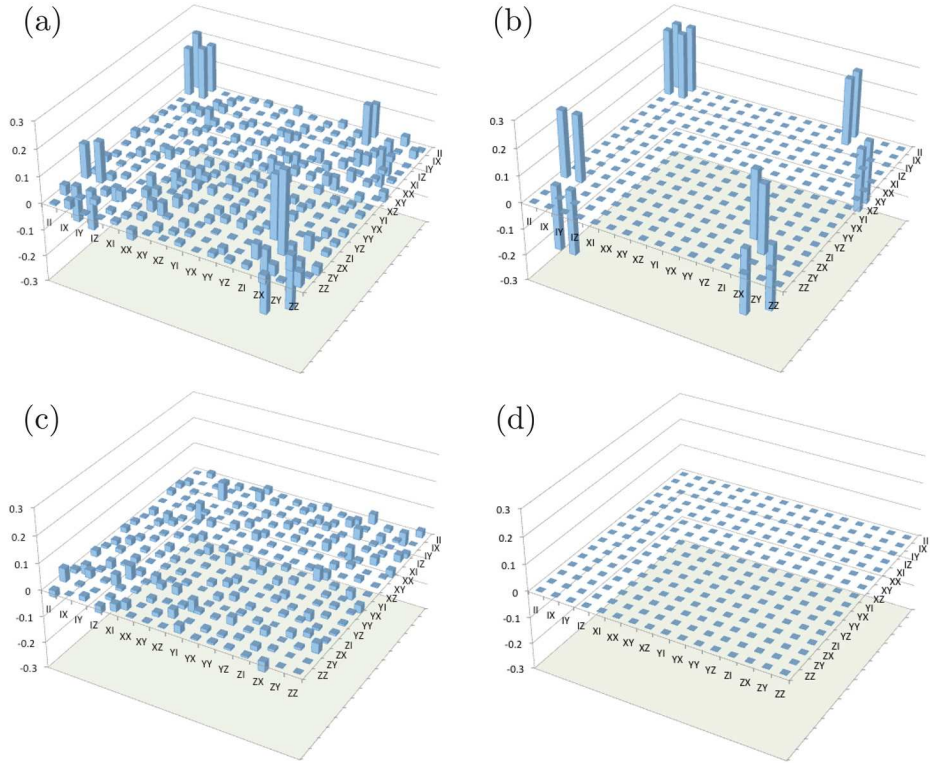}
\end{minipage}
\caption{{\sf CNOT} gate $\chi$ matrix. (a) Experimental real part. (b) Ideal real part. (c) Experimental imaginary part. (d) Ideal imaginary part.}
\label{cnot}
\end{figure*}

To perform the ${\sf T}$ gate requires three qubits in a linear cluster, as shown in Figure~\ref{gates}~(e), so here only qubit 1 is removed with a $\sigma_z$ measurement. The same encoding optics as before are used to set the input probe state on qubit 2 and state tomography is performed on qubit 4. Figure~\ref{tgate} shows the reconstructed experimental and ideal $\chi$ matrices. They appear similar, although the experimental case shows some unwanted probability of a $\sigma_x$ rotation, which results in bit-flip errors. Here, the process fidelity for the gate is found to be $F_{\sf T}=0.76 \pm0.04$. This fidelity is significantly higher than that of the ${\sf H}$ gate or the cluster state to the ideal cases, giving an error-probability per gate of $\epsilon > 0.24 \pm 0.04$. The fact that the ${\sf T}$ gate performs better than the ${\sf H}$ gate, which requires one less photon, suggests that the process of removing photons from the state with $\sigma_z$ measurements can cause the dephasing to have a greater effect on the gate. The result may be improved with smaller two- and three-photon cluster states directly used to carry out the ${\sf H}$ and ${\sf T}$ gates rather than starting from the four-photon state.

\subsection{{\sf CNOT} gate characterisation}

The {\sf CNOT} gate makes use of all four photons in the state as shown in Figure~\ref{gates}~(h), and because it is a two-qubit gate its characterisation requires all 16 combinations of $\ket{H}$,  $\ket{V}$,  $\ket{+}$, and  $\ket{L}$ across the two inputs. The encoding of the control qubit is performed on qubit 1 and the target on qubit 2, using the same techniques as before. Two-qubit state tomography is carried out on qubits 1 and 4 to measure the corresponding output for each encoding. The resulting $\chi$ matrices are shown in Figure~\ref{cnot}, using the joint Pauli basis for the 16 Kraus operators, $E_i \otimes E_j,~i,j=1,\dots 4$. The experimental and ideal cases appear similar, although there are many small background terms in the experimental case. The process fidelity is found to be $F_{\sf CNOT}=0.64 \pm 0.01$, giving an error-probability per gate of $\epsilon > 0.36 \pm 0.01$. This is higher than that of the ${\sf H}$ and ${\sf T}$ gates due to increased noise from the larger number of photons used and measurements required.

\section{Simulation of a Swap gate}
\begin{figure*}[t]
\begin{minipage}{13cm}
\hspace*{2.5cm}\includegraphics[width=12cm]{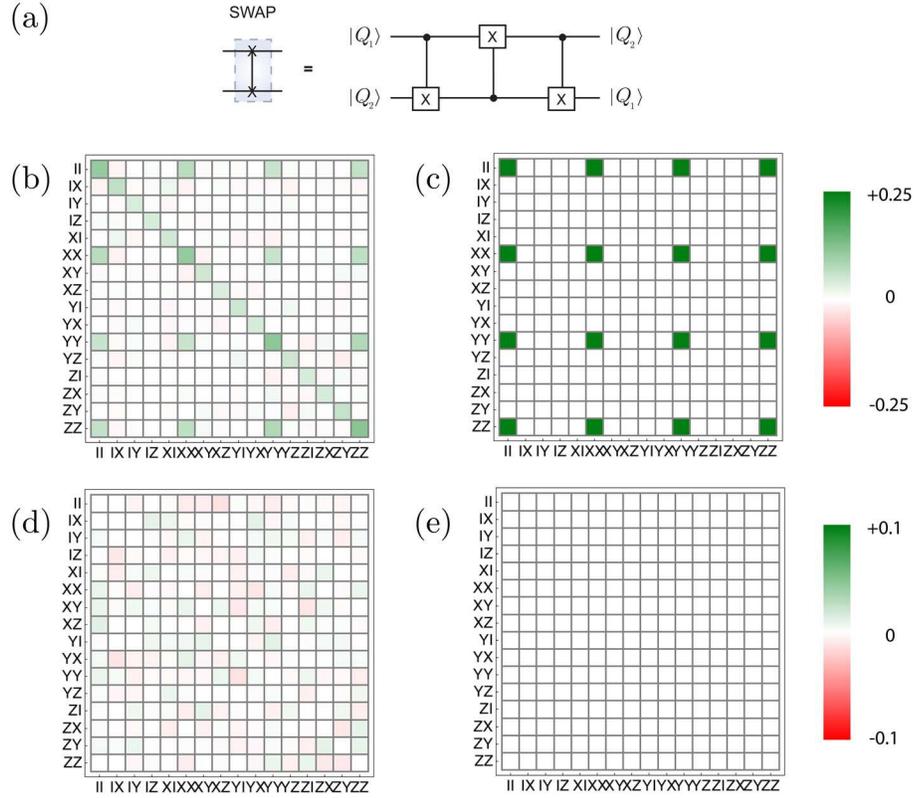}
\end{minipage}
\caption{{\sf SWAP} gate decomposition into universal gates (a) and simulated $\chi$ matrix: (b) Simulated real part. (c) Ideal real part. (d) Simulated imaginary part. (e) Ideal imaginary part. The unwanted diagonal elements in the simulated real part indicate depolarizing noise, and are about half the value ($\sim0.04$) of the desired diagonal and off-diagonal elements ($\sim0.08$)}
\label{Swap}
\end{figure*}

In order to give a basic idea of how our results can be used to gain an insight into the scalability of one-way quantum computing, we link up three {\sf CNOT} gates to simulate a swap gate~\cite{Barenco}, which we denote as {\sf SWAP}. This gate is another important building block for quantum computing as it allows the routing of information on a given cluster state resource for carrying out logic gates between logical qubits located far apart. The decomposition of the gate into {\sf CNOT}'s is shown in Figure~\ref{Swap}~(a) and the simulated $\chi$ matrix is shown in Figures~\ref{Swap}~(b) and (d), along with the ideal case in Figures~\ref{Swap}~(c) and (e). We find a simulated process fidelity of $F_{\sf SWAP}=0.30\pm 0.01$, giving an error-probability per gate of $\epsilon > 0.70 \pm 0.01$. One can see from Figure~\ref{Swap}~(b) that after concatenating only three {\sf CNOT} gates there is already a strong presence of diagonal components in the $\chi$ matrix which cause the simulated {\sf SWAP} gate to depart significantly from the ideal case. This suggests that the cumulative effect of imperfections in the individual gates is tending toward a local depolarizing channel on each qubit~\cite{NC}, which with some probability will leave them in a maximally mixed state rather than apply the correct operation. Note that here we have assumed the entangled resource for realising all the gates is generated from fused star cluster states, as shown in Figure~\ref{setup}, with ideal fusion. A more detailed analysis would also include the dephasing and multi-photon noise introduced during the fusion process, as described in Ref~\cite{Bell}. However, the current analysis can be viewed as giving an overall idea of the performance that can be expected, and the obvious need for improving the quality of the entangled photonic star cluster states, their fusion and the incorporation of error correction strategies~\cite{develop}. Although higher fidelity generation of cluster states and gate demonstrations do exist, to our knowledge none is close to reaching fault-tolerance, so that a similar build-up of errors will unavoidably occur as gates are concatenated. This highlights the need for scalable, low noise sources and manipulation of photonic entanglement.

%%%%%%%%%%%%%%%%%%%%%%%%%%%%%%%%%%
\section{Summary}

In this work we have generated a four-photon star cluster state capable of performing logic gates that are universal and efficient for quantum computing. By carrying out quantum process tomography on the CNOT, Hadamard, and T gates we are able to simulate any larger computation made up of combinations of these gates, which could include any quantum algorithm. As a basic example we simulated a Swap gate, finding that already with only three concatenated gates, the quality of the operation is significantly reduced. This is the first time these one-way gates have been characterized completely on a single resource state and our results provide important information about the requirements for building up more complex cluster states and for one-way quantum computation in realistic conditions. The limitations of the current experiment are largely due to the PCF source of entangled photons, in terms of multiphoton count rates and interference visibilities. With further development of the sources to improve the collection efficiency of the photons and their indistinguishability, it should become possible in future work to investigate experimentally larger cluster states and to perform a range of one-way protocols and algorithms.

%%%%%%%%%%%%%%%%%%%%%%%%%%%%%%%%%%
\ack

We acknowledge support from UK EPSRC, EU project 248095 Q-Essence, ERC grant 247462 QUOWSS, the Leverhulme Trust, and the Australian Research Council Centre of Excellence and DECRA schemes.  CUDOS is a Centre of Excellence (project number CE110001018).

%%%%%%%%%%%%%%%%%%%%%%%%%%%%%%%%%%
\appendix

%%%%%%%%%%%%%%%%%%%%%%%%%
%\appendix
%\section*{Appendix A: Derivation of fusion interference antidip}
%\setcounter{section}{1}

%%%%%%%%%%%%%%%%%%%%%%%%%
%\appendix
%\section*{Appendix B: Higher-order emission analysis}
%\setcounter{section}{2}


\begin{thebibliography}{99}
\bibitem{LS} The maximum likelihood approach is equivalent to a least-squares approach when the measurement errors are independent and normally distributed with constant standard deviation. For our counting events we have a Poissonian distribution, which converges to a Normal distribution for a large enough number of counts for each data point.
\end{thebibliography}
\end{document}